\documentclass[sigconf,natbib=true, review=False]{acmart} %
\usepackage{graphicx} %
\usepackage[show]{chato-notes}
\usepackage{amsmath}
\usepackage{algorithm}
\usepackage{subfig}
\usepackage{algpseudocode}
\usepackage{url}
\usepackage{makecell}
\usepackage{enumitem}
\usepackage{placeins}
\usepackage{multirow}
\usepackage{listings}
\usepackage{rotating}

\usepackage{marginnote}

\newcommand{\sasha}[1]{\textcolor[HTML]{000000}{#1}}
\newcommand{\rsasha}[1]{\textcolor[HTML]{000000}{#1}}
\newcommand{\gsasha}[1]{\textcolor[HTML]{000000}{#1}}

\newcommand{\nt}[1]{\textcolor{black}{#1}}
\newcommand{\pluseq}{\mathrel{+}=}

\newcommand{\scr}[1]{\textcolor[HTML]{000000}{#1}}

\newcommand{\scrc}[1]{\textcolor[HTML]{000000}{#1}}

\title{Efficient Inference of Sub-Item Id-based Sequential Recommendation Models with Millions of Items}
\copyrightyear{2024} 
\acmYear{2024}

\copyrightyear{2024}
\acmYear{2024}
\setcopyright{rightsretained}
\acmConference[RecSys '24]{18th ACM Conference on Recommender Systems}{October 14--18, 2024}{Bari, Italy}
\acmBooktitle{18th ACM Conference on Recommender Systems (RecSys '24), October 14--18, 2024, Bari, Italy}\acmDOI{10.1145/3640457.3688168}
\acmISBN{979-8-4007-0505-2/24/10}

\author{Aleksandr V. Petrov}
\affiliation{%
  \institution{University of Glasgow} \country{United Kingdom}}

\email{a.petrov.1@research.gla.ac.uk}

\author{Craig Macdonald}
\affiliation{%
  \institution{University of Glasgow} \country{United Kingdom}}
\email{craig.macdonald@glasgow.ac.uk}

\author{Nicola Tonellotto}
\affiliation{%
  \institution{University of Pisa} \country{Italy}}
\email{nicola.tonellotto@unipi.it}
\begin{document}

\begin{abstract}
Transformer-based recommender systems, such as BERT4Rec or SASRec, achieve state-of-the-art results in sequential recommendation. However, it is challenging to use these models in production environments with catalogues of millions of items: scaling Transformers beyond a few thousand items is problematic for several reasons, including high model memory consumption and slow inference.  %
In this respect, RecJPQ is a state-of-the-art method of reducing the \gsasha{models'} memory consumption; RecJPQ compresses item catalogues by decomposing item IDs into a small number of shared sub-item IDs. Despite reporting the reduction of memory consumption by a factor of up to $50\times$, the original RecJPQ paper did not report inference efficiency improvements over the baseline Transformer-based models. 
\gsasha{\scrc{Upon analysing} RecJPQ's scoring algorithm, we find that its efficiency is limited by \scrc{its use of score accumulators for each item}, which \scrc{prevents} parallelisation.} \rsasha{\scrc{In contrast}, LightRec (a non-sequential method that uses a similar idea of sub-ids) reported large inference efficiency improvements using an algorithm we call PQTopK.  We show that it is also possible to improve RecJPQ-based models' inference efficiency using the PQTopK algorithm. } In particular, we speed up RecJPQ-enhanced SASRec by a factor of 4.5$\times$ compared to the original SASRec's inference method and by the factor of  1.56$\times$ compared to the method implemented in RecJPQ code on a large-scale Gowalla dataset with more than million items. Further, using simulated data, we show that PQTopK remains efficient with catalogues of up to tens of millions of items, removing one of the last obstacles to using Transformer-based models in production environments with large catalogues.
\end{abstract}

\maketitle

\section{Introduction} \label{sec:intro}
The goal of sequential recommender models is to predict the next item in a sequence of user-item interactions. The best models for this task, such as SASRec~\cite{SASRec} and BERT4Rec~\cite{BERT4Rec}, are based on adaptations of the Transformer~\cite{Transformer} architecture. 
Indeed, while the \scrc{Transformer} architecture was originally developed for natural language processing, sequential recommender systems adapt the architecture by using tokens to represent items, \rsasha{and \gsasha{the} next item prediction task then becomes equivalent to the next token prediction task in the language models.}

\looseness -1 Despite achieving state-of-the-art results on \rsasha{datasets available in academia}, it is challenging to use these models in a production environment due to the scalability issues: the number of items in large-scale recommender systems, such as product recommendations in Amazon, can reach hundreds of millions~\cite{AmazonStatisticsUptoDate}, which is much larger than tens of thousands of tokens in the dictionaries of language models. A large catalogue causes \scr{several problems} in Transformer models, such as large GPU memory requirements to store the item embeddings \rsasha{during training}, large computational resources required to train models, and slow inference in production. Several works have recently addressed the memory consumption issues~\cite{xiaEfficientOnDeviceSessionBased2023, petrovRecJPQTrainingLargeCatalogue2024} and inefficient training~\cite{klenitskiyTurningDrossGold2023, petrovGSASRecReducingOverconfidence2023, petrovRSSEffectiveEfficient2025}; however, efficient model inference remains an open question, which is the focus of this paper.

\rsasha{Efficient inference is especially important when considering a model deployment on CPU-only hardware (i.e. without GPU acceleration). Indeed, deploying a trained model on CPU-only hardware is often a practical choice, considering the high \rsasha{running} costs associated with GPU accelerators. 
Hence, in this paper, we specifically focus on the CPU-only inference efficiency of Transformer-based sequential recommendation models.}

\begin{figure*}[tb]\hspace{-3mm}
    \vspace{-0.5\baselineskip}
\includegraphics[width=0.8\textwidth]{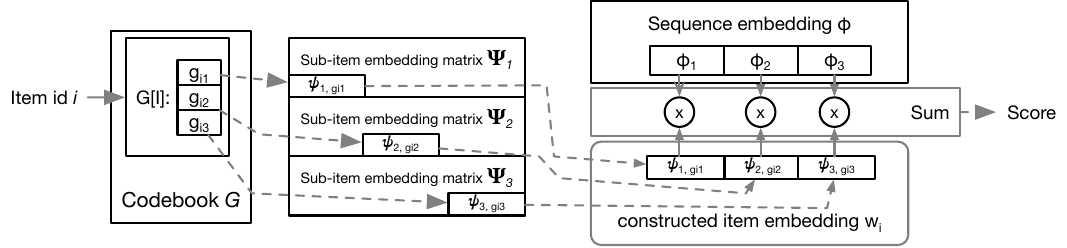}%
    \caption{Reconstruction of item embeddings \sasha{and computing item scores} using Product Quantisation \sasha{with $m=3$ splits.} }\vspace{-.75\baselineskip}
    \label{fig:embedding_reconstruction}
\end{figure*}

\looseness -1 \scrc{The inference of a Transformer-based recommendation model consists of two parts: computing a {em sequence representation} using the \emph{backbone} Transformer model, followed by computing the scores of individual items using this representation (see Section~\ref{ssec:recess:preliminarilies} for details).} The main cause of the slow inference by Transformer-based models arises not from the Transformer backbone model itself but from the computation of \scrc{all the item scores}. Indeed, the inference time of a given Transformer backbone model is constant w.r.t.\ the number of items \rsasha{(after embedding lookup, which is $\mathcal{O}(1)$ operation, the Transformer model only works with embeddings, which do not depend on the number of items)}; however, computing item scores has a linear complexity w.r.t.\ the number of items. Hence, to speed up inference, there are three options: (i) reduce the number of scored items, (ii) reduce the number of operations per item, and (iii) efficiently parallelise computations. 
In the first category are the approximate nearest neighbour methods, such as FAISS~\cite{FAISS} or Annoy~\cite{SpotifyAnnoy2024}. While these methods can be practical in some cases, there are two problems: (i) these methods are \scrc{\emph{unsafe}~\cite{turtleQueryEvaluationStrategies1995b,tonellottoEfficientQueryProcessing2018}}, meaning that the results retrieved using an ANN index may omit some candidates that would have been scored high by the model and (ii) they require item embeddings to be present in the first place in order to build the index, and training item embeddings for all items in large catalogue case may not be feasible in the first place~\cite{petrovRecJPQTrainingLargeCatalogue2024}.
Therefore, this paper focuses on \scr{analysing the efficiency of existing methods and} reducing the number of operations per item and parallelising the computations. In particular, we build upon RecJPQ~\cite{petrovRecJPQTrainingLargeCatalogue2024}, a recent state-of-the-art approach for compressing embedding tables in Transformer-based sequential recommenders. RecJPQ achieves compression by representing items using a concatenation of shared sub-ids. While achieving great results on compression (for example, on the Gowalla~\cite{choFriendshipMobilityUser2011} dataset, RecJPQ achieves up to 50$\times$ compression without degrading effectiveness), the RecJPQ paper does not perform any analysis of the model's inference in large catalogues and only briefly mentions that it is similar to the inference time of the original non-compressed models. On the other hand, prior works that built upon similar ideas of sub-id-based recommendation, such as LightRec~\cite{lianLightRecMemorySearchEfficient2020}, showed that the sub-id-based method could indeed improve model inference time. \rsasha{Inspired by LightRec, we describe a sub-id-based scoring algorithm for PQ-based models, which we call \textit{PQTopK}.} \scr{We further} analyse if RecJPQ-enhanced Transformer-based recommendation models can be efficiently inferred on catalogues with (multiple) millions of items using the PQTopK algorithm in a CPU-only environment.

\looseness -1 The main contributions of this paper can be summarised as follows: (i) we analyse inference efficiency of RecJPQ-enhanced versions of SASRec~\cite{SASRec} and BERT4Rec~\cite{BERT4Rec} and find that it is more efficient than Matrix-Multiplication based scoring used in the original models; (ii) we show that scoring \gsasha{efficiency} of RecJPQ-based models can be \gsasha{improved using the PQTopK algorithm} (iii) we explore the limits of \gsasha{PQTopK}-based inference using simulated settings with up to 1 billion items in catalogue and show that inference remains efficient with millions of items.
To the best of our knowledge, this is the first analysis of the inference of sub-id-based sequential models on large-scale datasets and the first demonstration of the feasibility of using these models in the large-catalogue scenario.

\vspace{-1\baselineskip}
\section{Background}\label{ssec:recess:preliminarilies}
\paragraph{Large-catalogue Sequential RecSys}\label{sec:rec}\
\looseness -1 Usually, sequential recommendation is cast as the \emph{next item prediction} task. Formally, given a sequence of user-item interactions $\left\{i_1, i_2, i_3 ... i_n\right\}$, \nt{also known as their \textit{interactions history},} the goal of a recommender system is to predict the next item in the sequence, $i_{n+1}$ from the \emph{item catalogue} (the set of all possible items) $I$. The total number of items $|I|$ is the \emph{catalogue size}.

\looseness -1 Typically, to generate recommendations given a history of interactions $h = \left\{i_1, i_2, i_3 ... i_n\right\}$, Transformer-based models first generate a sequence embedding $\phi \in \mathbb{R}^d$.
The scores for all items, $r = (r_1, \ldots, r_{|I|}) \in \mathbb{R}^{|I|}$, are then computed by multiplying the matrix of item embeddings $W \in \mathbb{R}^{|I|\times d}$, which is usually shared with the embeddings layer of the Transformer model, by the sequence embedding $\phi$, i.e., $r = W\phi$.
Finally, the model generates recommendations by selecting \gsasha{from} $r$ the top $K$ items with the highest scores.%

Despite their effectiveness, training Transformer-based models with large item catalogues is a challenging task, as these models typically have to be trained for long time~\cite{Bert4RecRepro} and require appropriate selection of training objective~\cite{petrovRSSEffectiveEfficient2025}, negative sampling strategy and loss function~\cite{petrovGSASRecReducingOverconfidence2023,klenitskiyTurningDrossGold2023}.
Transformer-based models with large catalogues also require a lot of memory to store the item embeddings $W$. This problem has recently been addressed by Product Quantisation, which we describe in the next section. 
Finally, another problem with Transformer-based models with large catalogues is their slow inference with large catalogues. Indeed, computing all item scores 
may be prohibitively expensive when the item catalogue is large: it requires $|I|\times d$ scalar multiplications and additions, and, as we noted in Section~\ref{sec:intro},  in real-world recommender systems, the catalogue size $|I|$ may reach hundreds of millions of items, making exhaustive computation 
impractical. Moreover, typically large-catalogue recommender systems have a large number of users as well; therefore, the model has to be used for inference very frequently and, ideally, using only low-cost hardware, i.e., without using GPU acceleration. Therefore, real-life recommender systems rarely exhaustively score all items for all users and instead apply unsafe heuristics (i.e.\ heuristics that do not provide theoretical guarantees that all high-scored items will be returned, such as two-stage ranking). 
 
 \vspace{-0.5\baselineskip}

\looseness -1 \paragraph{Sub-Id based recommendation}\label{ssec:recjpq} When the catalogue size reaches several million items, the item embedding matrix $W$ becomes too large to fit in a GPU's memory~\cite{petrovRecJPQTrainingLargeCatalogue2024}. To reduce its memory footprint, several recent recommendation \scrc{models~\cite{petrovRecJPQTrainingLargeCatalogue2024, liuVectorQuantizationRecommender2024}} adopted Product Quantisation (PQ)~\cite{jegouProductQuantizationNearest2011} -- a well-studied method for vector compression. PQ is a generic method that can be applied to any \nt{set of vectors}; however, in this section, we \nt{describe} PQ as applied to item embeddings. 

\looseness -1 To compress the item embedding matrix $W$, PQ associates each item id to a list of {\em sub-ids}, akin to language models breaking down words into sub-word tokens. PQ reconstructs an item's embedding by combining the sub-id embeddings assigned to it. These sub-ids are organised into {\em splits}, with each item having precisely one sub-id from each split\footnote{We use the terminology from~\cite{petrovRecJPQTrainingLargeCatalogue2024}. In the PQ literature, another popular term for sub-id embeddings is \emph{centroids}, and another popular term for splits is \emph{sub-spaces}.}. More formally, PQ first builds a \emph{codebook} $G\in\mathbb{N}^{|I|\times m}$ that maps an item id $i$ to its associated $m$ sub-ids:
\begin{align}
    G[i] \rightarrow \left\{g_{i1}, g_{i2}, ..., g_{im}\right\} \label{eq:sub_ids_map}
\end{align}
where $g_{ij}$ is the $j$-th sub-id associated with item $i$ and $m$ is the number of splits. There are several algorithms for associating ids to sub-ids, for instance, using KMeans~\cite{jegouProductQuantizationNearest2011}, or learning during model training~\cite{chenDifferentiableProductQuantization2020}; 
RecJPQ~\cite{petrovRecJPQTrainingLargeCatalogue2024} derives the codes from a truncated SVD decomposition of the user-item interactions matrix. 

For each split $k=1,\ldots,m$, PQ stores a sub-item embedding matrix $\Psi_k \in \mathbb{R}^{b\times\frac{d}{m}}$, where $b$ is the number of distinct sub-ids in each split. The $j$-th row of $\Psi_k$ denotes the sub-item embedding $\psi_{k,j} \in \mathbb{R}^{\frac{d}{m}}$ associated with the $j$-th sub-id, in the $k$-th split.
Then, PQ reconstructs the item embedding $w_i$ as a concatenation of the associated sub-id embedding: 
\begin{align}
    w_i =  \psi_{1,g_{i1}} \mathbin\Vert \psi_{2,g_{i2}}  \mathbin\Vert ... \mathbin\Vert  \psi_{m,g_{im}} \label{eq:pq:item_embedding}
\end{align}

Finally, an item score is computed as the dot product of the sequence embedding and the constructed item embedding: 
\begin{align}
    r_i = w_i \cdot \phi\label{eq:dot_product}
\end{align}
\looseness -1 A straightforward use of Equation~\eqref{eq:dot_product} for item scoring in PQ-based recommendation models does not lead to any computational efficiency improvements compared to models where all item embeddings are stored explicitly: \rsasha{in both cases, the algorithm would need to multiply the sequence embedding $w$  by the (reconstructed) embeddings of all items.}
However, \rsasha{the sub-id representations of} PQ \rsasha{allow} a more efficient scoring algorithm, \rsasha{which we describe next}.

\vspace{-0.5\baselineskip}
\section{PQTopK Algorithm} \label{sec:pq_topk}
\rsasha{PQTopK is a scoring algorithm for PQ-based models \rsasha{that uses pre-computation of sub-id scores for improved inference efficiency}. While versions of this algorithm have previously been described, for example, for a different recommendation scenario~\cite{lianLightRecMemorySearchEfficient2020} and for document retrieval~\cite{zhanJointlyOptimizingQuery2021}, to the best of our knowledge, it has not been previously applied for sequential recommendation nor Trans\-former-based models.}

\looseness -1 PQTopK first splits the sequence embedding $\phi \in \mathbb{R}^{d}$ \rsasha{obtained from a Transformer model} into $m$ sub-embeddings $\left\{\phi_1, \phi_2 ... \phi_m\right\}$, with $\phi_{k} \in \mathbb{R}^{\frac{d}{m}}$ for $k=1, \ldots,m$, such that $\phi = \phi_1  \mathbin\Vert \phi_2  \mathbin\Vert ... \mathbin\Vert  \phi_m$.
By substituting Equation~\eqref{eq:pq:item_embedding} and the similarly decomposed sequence embedding $\phi$ into Equation~\eqref{eq:dot_product}, the final item score for item $i$ is obtained as the sum of sub-embedding dot-products: 
\begin{align}
   r_i = w_i \cdot \phi &= (\psi_{1,g_{i1}} \mathbin\Vert ... \mathbin\Vert  \psi_{m,g_{im}}) \cdot (\phi_1  \mathbin\Vert ... \mathbin\Vert  \phi_m) = \sum_{k=1}^m \psi_{k,g_{ik}} \cdot \phi_k \nonumber
\end{align}

Let $S \in \mathbb{R}^{m \times b}$ denote the \emph{sub-id score matrix}, which consists of \emph{sub-id scores} $s_{k,j}$, defined as dot products of the sub-item embedding $\psi_{k,j}$ and the sequence sub-embeddings $\phi_k$:
\begin{align}
s_{k,j} = \psi_{k,j} \cdot \phi_k\label{eq:sub_item_scores}
\end{align}

The final score of item $i$ can, therefore, also be computed as the sum of the scores of its associated sub-ids:
\begin{align}
   r_{i} = \sum_{k=1}^m s_{k,g_{ik}} \label{eq:sum_sub_scores}
\end{align}
\scrc{Figure~\ref{fig:embedding_reconstruction} also graphically illustrates how item scores are computed using PQ.}

\looseness -1 The number of splits $m$ and the number of sub-ids per split $b$ are usually chosen to be relatively small, 
so that the total number of sub-id scores is much less compared to the size of the catalogue, e.g., $m\times b \ll |I|$.
Therefore, \scrc{this allows to compute the matrix $S$ only once for a given sequence embedding} and then reuse these scores for all items. This leads to efficiency gains compared to matrix multiplication, as scoring each item now only requires $m \ll d$ additions instead of $d$ multiplications and $d$ additions per item. %
The time for pre-computing sub-item scores %
does not depend on $|I|$ %
and 
we can assume that it is negligible w.r.t.\ the exhaustive scoring of all items. 

\begin{algorithm}[tb]
\small
\caption{PQTopK($G$, $S$, $K$, $V$).}\label{alg:top_k}
\begin{algorithmic}[1]
   \Require $G$ is the codebook (mapping: item id $\rightarrow$ sub-item ids), Eq.~\eqref{eq:sub_ids_map}
   \Require $S$ is the matrix of pre-computed sub-item scores, indexed by split and sub-item, Eq.~\eqref{eq:sub_item_scores}

   \Require $K$ is the number of results to return
   \Require $V \subseteq I$ are the items to score; all items ($V = I$)  if not given 
   
   \State $scores \gets$ empty array of scores for all items in $V$, initialised to 0
   \For{$item\_id \in V$} \label{alg:outer_loop} \Comment{This loop can be efficiently parallelised}
        \State $score[item\_id] \gets \sum_{k=1}^{m} S[k,G[item\_id,k]] \label{alg:inner_loop} $ \Comment{Eq.~\eqref{eq:sum_sub_scores}}
   \EndFor
   \State \Return TopK($score$, $K$) \Comment{Returns a list of $\langle$ItemId, Score$\rangle$ pairs} 
\end{algorithmic}
\end{algorithm}

Algorithm~\ref{alg:top_k} illustrates the PQTopK in pseudo-code. Note that the algorithm has two loops: the outer loop (line~\ref{alg:outer_loop}) iterates over the items in the catalogue, and the inner loop (line~\ref{alg:inner_loop}) iterates over codes associated with the item. However, as the item scores are independent of each other, both loops can be efficiently parallelised\footnote{\scrc{We achieve parallelisation using Tensorflow accelerated computation framework.}}. 

\looseness -1 The original RecJPQ~\cite{petrovRecJPQTrainingLargeCatalogue2024} code is also based on the same idea of pre-computing item scores and then computing item scores as the sum of associated sub-id scores. However, in RecJPQ, the order of loops is swapped compared to the PQTopK algorithm: the outer loop iterates over the splits, and in the inner loop, the scores for each item are accumulated \scrc{for each item} (\scrc{we list RecJPQ's original scoring algorithm in  Algorithm~\ref{alg:RecJPQtop_k}}). Due to the \scrc{iterative accumulation of item scores}, the outer loop in RecJPQ's scoring algorithm is not parallelised. \scrc{In Section~\ref{sec:results}, we show} that this makes RecJPQ's scoring algorithm less efficient compared to PQTopK. 

\begin{algorithm}[h]
\small
\caption{RecJPQScore($G$, $S$, $K$, $V$) Scoring algorithm \scrc{originally} used in RecJPQ.}\label{alg:RecJPQtop_k}
\begin{algorithmic}[1]
   \Require $G$ is the codebook (mapping: item id $\rightarrow$ sub-item ids), Eq.~\eqref{eq:sub_ids_map}
   \Require $S$ is the matrix of pre-computed sub-item scores, indexed by split and sub-item, Eq.~\eqref{eq:sub_item_scores}

   \Require $K$ is the number of results to return
   \Require $V \subseteq I$ are the items to score; all items ($V = I$)  if not given 
   
   \State $scores \gets$ empty array of scores for all items in $V$, initialised to 0
   \For{$k \in 1..m$} \Comment{Not parallelised in RecJPQ}
        \For{$item\_id \in V$} 
            \State $score[item\_id]  \pluseq S[k,G[item\_id,k]] $ 
        \EndFor
   \EndFor
   \State \Return TopK($score$, $K$) 
\end{algorithmic}
\end{algorithm}

\FloatBarrier
\vspace{-0.5\baselineskip}
\section{Experimental Setup}\label{sec:expsetup}
We designed our experiments to answer two research questions: 
\begin{itemize}
    \item[RQ1] How does PQTopK inference efficiency compare to baseline item scoring methods? 
    \item[RQ2] How does PQTopK inference efficiency change when increasing the number of items in the catalogue? 
\end{itemize}

\begin{table}[t]
\caption{Salient characteristics of the experimental datasets.} \label{tb:datasets}
    \resizebox{\linewidth}{!}{
      
    \begin{tabular}{lrrrrr}
    \toprule
    Dataset &  Users &  Items &  Interactions &  {Avg. length} \\
    \midrule
    Booking.com &   140,746 &      34,742 &           917,729 &            6.52 \\
    Gowalla  &      86,168 &    1,271,638 &           6,397,903 &            74.24 \\
    \bottomrule
    \end{tabular}

}
\end{table}

\textit{Datasets.} We experiment with two real-world datasets: Booking.com~\cite{goldenberg_bookingcom_2021}  ($\sim$35K items) and Gowalla~\cite{choFriendshipMobilityUser2011} ($\sim$1.3M items). Following common practice, we remove users with less than five items from the data. Salient characteristics of the experimental data are provided in Table~\ref{tb:datasets}. Additionally, to test the inference speed of different scoring methods, we use simulated data with up to 1 billion items in the catalogue.

\looseness -1 \textit{Backbone Models.} \scr{In RQ1, } we experiment with two commonly used Transformer models: SASRec and BERT4Rec. To be able to train the models on large catalogues, we replace the item embedding layer with RecJPQ~\cite{petrovRecJPQTrainingLargeCatalogue2024}. Moreover, the original BERT4Rec does not use negative sampling, which makes it infeasible to train on large catalogues, such as Gowalla. Hence, to be able to deal with large catalogues, we use gBERT4Rec~\cite{petrovGSASRecReducingOverconfidence2023}, a version of BERT4Rec trained with negative sampling and gBCE loss. The configuration of the models follows the details described in the RecJPQ paper~\cite{petrovRecJPQTrainingLargeCatalogue2024}. In particular, we use 512-dimensional embeddings;  we use 2 Transformer blocks for SASRec and 3 Transformer blocks for BERT4Rec. When answering RQ1, we use RecJPQ with $m=8$ splits but vary $m$ in RQ2. \scr{\scrc{In RQ2, we exclude the backbone model from our analysis;} therefore, \scrc{the results are model-agnostic and apply to any backbone}.}

\looseness -1 \textit{Scoring Methods.} We analyse three scoring methods: (i) Transformer Default,  matrix multiplication-based scoring $r = W\phi$ %
used by default in SASRec and BERT4Rec \rsasha{(w/o any PQ enhancements)}; (ii) the original RecJPQ \scrc{scoring} \scrc{(Algorithm~\ref{alg:RecJPQtop_k})}; (iv) PQTopK scoring (\scr{Algorithm~\ref{alg:top_k}}). We implement\footnote{\scrc{Code for the paper: \href{https://github.com/asash/RecJPQ-TopK}{https://github.com/asash/RecJPQ-TopK}}.} all algorithms using TensorFlow~\cite{abadiTensorFlowSystemLargeScale2016}. 

\textit{Metrics.} Our main focus is on the model inference speed. We measure inference using the median response time \scr{per user} (mRT, time required by the model to return recommendations). We do not use GPU acceleration when measuring any response time \rsasha{(details of our hardware configuration are in Table~\ref{tb:hardware})}. We separately measure total response time, time spent by the model for running the backbone Transformer model, and time spent by the scoring algorithm. For completeness, we also report effectiveness using NDCG@10, even though optimising model effectiveness is outside of the scope of the paper \scrc{and all scoring methods for RecJPQ-based models have the same effectiveness}.

\vspace{-0.5\baselineskip}
\section{Analysis}\label{sec:results}

\looseness -1 \textit{RQ1. Comparison of PQTopK and other scoring methods.} Table~\ref{tb:main} reports effectiveness and efficiency metrics for SASRec and BERT4\-Rec on both Booking.com and Gowalla datasets. We first observe that nDCG@10 values do not depend on the scoring method, as all algorithms compute the same score distribution. We also see that \rsasha{the model backbone model inference time} does not depend on the scoring method as well, as different scoring methods are applied on top of the backbone Transformer model (i.e.\ we use different ``heads" in Transformer terminology). Interestingly, the time required by the backbone Transformer model does not depend on the dataset either: e.g., BERT4Rec requires roughly 37 milliseconds on both Booking and Gowalla, while SASRec requires roughly 24 milliseconds. This makes sense as Transformer complexity depends on the embedding dimensionality, the number of Transformer blocks and the sequence length but not on the number of items in the catalogue.

On the smaller Booking.com dataset, we see that \scrc{the running time of the backbone Transformer model dominates the total model response time}, and the differences between different scoring methods are rather unimportant. For example, when using gBERT4Rec on this dataset, the slowest scoring method (Transformer Default) requires 43 milliseconds per user. In contrast, the faster method (PQTopK) requires 40 milliseconds ($\Delta$<10\%) -- \rsasha{even though PQTopK is two times faster compared to Transformer Default scoring when comparing without the backbone model inference.}  In contrast, on the larger Gowalla dataset with more than 1M items, there is a large difference between different scoring methods. For example, when using Default Transformer scoring with SASRec, inference time is dominated by the item scoring (131ms out of 171ms). 

\begin{table}[tb]
    \centering
    \caption{Hardware Configuration}
    \begin{tabular}{c|c}
        \toprule
         CPU & AMD Ryzen 5950x  \\
         Memory & 128 GB DDR4 \\ 
         OS & Ubuntu 22.04.3 LTS  \\
         Accelerated computing framework & TensorFlow 2.11.0 \\
         GPU Acceleration & Not used\\
         \bottomrule
    \end{tabular}
    \label{tb:hardware}
\end{table}

\begin{table*}
\caption{Efficiency analysis of item scoring methods. mRT is the Median Response Time, measured in milliseconds; SAS is the SASRec model and BERT is the gBERT4Rec model.} \label{tb:main}
\vspace{-1\baselineskip}
    \begin{tabular}{|l|l|cccccc|}
    \hline
                               &                     & \multicolumn{3}{c|}{Dataset: Booking}                                                                                                                                                                                                  & \multicolumn{3}{c|}{Dataset: Gowalla}                                                                                                                                                                            \\ \hline
                   &\makecell[l]{Scoring\\method}      & \begin{tabular}[c]{@{}c@{}}mRT\\ (Scoring)\end{tabular} & \begin{tabular}[c]{@{}c@{}}mRT\\ (Total)\end{tabular} & \multicolumn{1}{c|}{\begin{tabular}[c]{@{}c@{}}Backbone\\ measures\end{tabular}}                                    & \begin{tabular}[c]{@{}c@{}}mRT\\ (Scoring)\end{tabular} & \begin{tabular}[c]{@{}c@{}}mRT\\ (Total)\end{tabular} & \begin{tabular}[c]{@{}l@{}}Backbone\\ measures\end{tabular}                                    \\ \hline
    \multirow{3}{*}{\rotatebox{90}{BERT}} & Default & 6.22                                                    & 43.37                                                 & \multicolumn{1}{l|}{\multirow{3}{*}{\begin{tabular}[c]{@{}c@{}}NDCG@10:  0.328 \\ Model mRT: \\37.16\end{tabular}}}  & 133.40                                                  & 171.04                                                & \multirow{3}{*}{\begin{tabular}[c]{@{}c@{}}NDCG@10:  0.168\\  Model mRT: \\ 37.52\end{tabular}} \\
                               & RecJPQ     & 3.90                                                     & 41.08                                                 & \multicolumn{1}{l|}{}                                                                                               & 33.87                                                   & 71.42                                                 &                                                                                                \\
                               & \gsasha{PQTopK}             & \textbf{3.09}                                           & \textbf{40.23}                                        & \multicolumn{1}{l|}{}                                                                                               & \textbf{13.79}                                          & \textbf{51.33}                                        &                                                                                                \\ \hline
    \multirow{3}{*}{\rotatebox{90}{SAS}}    & Default & 6.27                                                    & 30.03                                                 & \multicolumn{1}{l|}{\multirow{3}{*}{\begin{tabular}[c]{@{}c@{}}NDCG@10: 0.188\\  Model mRT:\\ 23.75\end{tabular}}} & 131.35                                                  & 156.07                                                & \multirow{3}{*}{\begin{tabular}[c]{@{}c@{}}NDCG@10: 0.120\\  Model mRT: \\ 24.67\end{tabular}} \\
                               & RecJPQ     & 3.77                                                    & 27.53                                                 & \multicolumn{1}{l|}{}                                                                                               & 29.65                                                   & 54.32                                                 &                                                                                                \\
                               & \gsasha{PQTopK}             & \textbf{2.93}                                           & \textbf{26.69}                                        & \multicolumn{1}{l|}{}                                                                                               & \textbf{10.03}                                          & \textbf{34.72}                                        &                                                                                                \\ \hline
    \end{tabular}
\end{table*}

\looseness -1 When using  \gsasha{SASRec as the backbone with RecJPQ scoring}, both \gsasha{the backbone and the scoring head} contribute \gsasha{similarly} towards total scoring time (SASRec takes 24ms while scoring takes 29ms). \rsasha{In contrast}, when using PQTopK, the total time \rsasha{is dominated by} the Transformer model itself (e.g., PQTopK only uses 10\gsasha{ms.}  out of 34 when using the SASRec backbone). If we isolate scoring time, the Gowalla with SASRec backbone dataset with  PQTopK is 13$\times$ faster than the Transformer default and $3\times$ faster than RecJPQ scoring.

In summary, answering RQ1, we find that PQTopK is the most efficient method among the baselines. \rsasha{On the Gowalla dataset with more than a million items, PQTopK} requires much less time compared to backbone Transformer models. On the other hand, on smaller datasets with only a few thousand items (such as Booking.com), even Default Matrix Multiplication remains efficient.

\textit{RQ2. PQTopK efficiency with very large catalogues.} As observed in RQ1, the inference time of a backbone Transformer model (without scoring head) is constant \rsasha{w.r.t.\ catalogue size $|I|$}. Therefore, as our goal is efficiency analysis, we exclude the Transformer model from the analysis and simulate it using a random model output for each output. \scr{We also generate} a random sub-id embedding matrix $\Psi$ to compute item scores. In all cases, we include the time required for selecting top-k (\rsasha{\texttt{tf.math.top\_k()} in TensorFlow}) after scoring, as this time also depends on the number of items in the catalogue.

\begin{figure}
    \vspace{-1\baselineskip}
    \centering
        \subfloat[Number of splits: $m=8$]{
        \resizebox{0.7\linewidth}{!}{
            \includegraphics{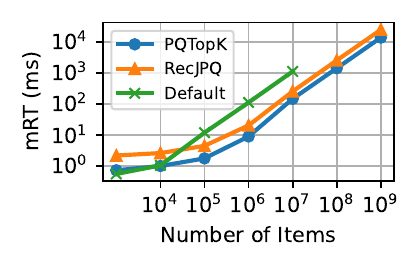}
            \label{subfig:simpulated_8}
        }
        }\\
        \subfloat[Number of splits: $m=64$]{
        \resizebox{0.7\linewidth}{!}{
            \includegraphics{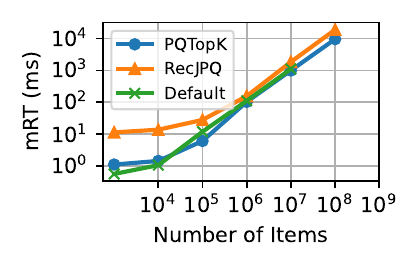}
            \label{subfig:simpulated_64}
        }
        }
    \vspace{-0.8\baselineskip    }
    \caption{Efficiency of PQTopK on simulated data}
    \label{fig:simulate_effectiveness}
\end{figure}

Figure~\ref{fig:simulate_effectiveness} reports the mean response time for Default Transformer scoring, PQTopK and RecJPQ without the backbone Transformer model, for  $m=8$ splits (\ref{subfig:simpulated_8}) and $m=64$ splits (\ref{subfig:simpulated_64}).
Both \scr{Figures}~\ref{subfig:simpulated_8} \&~\ref{subfig:simpulated_64} include the matrix multiplication-based Transformer Default baseline that does not \gsasha{use} the number of splits. 

We observe from the figures that with \scr{a} a low number of items in the catalogue ($\leq 10^4$), the default matrix multiplication-based approach is the most efficient, requiring less than a millisecond for scoring. However, as we observed in RQ1, with this small number of items, the actual method is not that important, as the scoring time is likely to be dominated by the backbone model inference. 

With the smaller number of splits, $m=8$, matrix multiplication becomes \gsasha{less efficient compared to PQ-based methods for item catalogues with more than $10^5$} items. Note that the figure is shown in logarithmic scale, meaning that, for example, at 10M items, PQTopK is \gsasha{10$\times$} more efficient compared to the default approach. Also, note that the matrix multiplication baseline only extends up to $10^7$ items: after that point, the default approach exhausts all available system memory (128GB). We also observe that PQTopK is always more efficient than RecJPQ. Despite (due to the logarithmic scale) the lines looking close to each other, PQTopK is always faster than RecJPQ by 50-100\%. For example, with 10M items in the catalogue, PQTopK requires 146ms per user, whereas \scr{RecJPQ} requires 253ms (+68\%). With 100M items in the catalogue, PQTopK remains relatively efficient ($\approx$ 1 second per user); however, with 1 billion items, the method requires more than 10 seconds \scr{per user}. \rsasha{Aruguably, 10 seconds per item is not suitable for interactive recommendations (for example, when the model inference occurs during web page loading), but may still \gsasha{work} in \gsasha{suit} situations when recommendations can be pre-computed \gsasha{(e.g. updated once every day)}. }

\looseness -1 On the other hand, as we can see from Figure~\ref{subfig:simpulated_64}, \gsasha{with} a large number of splits ($m=64$), \gsasha{Default} and PQtopK perform \gsasha{similarly}; \gsasha{e.g.}, both methods require  $\sim$100ms for scoring 1M items, \rsasha{50ms faster than RecJPQ}. However, \scrc{on our hardware,}  \gsasha{Default} consumes all available memory above 10M items \scrc{(this is why the line for Default on Figures~\ref{subfig:simpulated_64} and~\ref{subfig:simpulated_64} does not go beyond $10^7$ items)}, whereas PQTopK and RecJPQ allow for scores up to \scrc{100M items}. \scrc{Nevertheless}, PQTopK scoring, in this case, requires 10 seconds per user, limiting its application to the pre-computing scenario. 

\looseness -1 \gsasha{Finally, we observe that with catalogues with more than  $10^5$ items, the response depends linearly on the number of items for all scoring methods}. \gsasha{However, with less than $10^5$ items, there is a }  "elbow-style" non-linearity \gsasha{that} can be explained \gsasha{by the fact that the time required} by auxiliary operations such as function calls becomes important at this small scale. 

\looseness -1 \gsasha{Summarising} RQ2, we \gsasha{conclude} that PQTopK with 8 splits is a very efficient algorithm that allows performing efficient inference on catalogues even with hundreds of millions of items. With a larger number of splits $m=64$, the inference time of PQTopK is similar to the default matrix multiplication scoring, but it allows scoring up to $10^8$ items. In contrast, matrix multiplication \gsasha{exhausts available memory} with catalogues larger than  $10^7$ items, \gsasha{which highlights the importance of RecJPQ for reducing memory consumption}. 

\section{Conclusion}\label{sec:conclusion}
\looseness -1 This paper analysed the inference time of Transformer-based sequential recommender systems with large catalogues. \gsasha{We found that using RecJPQ enhancement, which enables training on large catalogues via sub-item-id representation, coupled with an efficient PQTopK scoring algorithm,} allows model inference on large catalogues. In particular, \gsasha{using PQTopK, we sped up RecJPQ-enhanced SASRec 1.56$\times$ compared to the original RecJPQ scoring and 4.5$\times$ compared to default SASRec scoring} \rsasha{on the Gowalla dataset with 1.3M items}. We also showed that, when considering the pre-scoring scenario, PQTopK can be applicable to catalogues of up to 1 billion items. We believe that our findings will help the wider adoption of state-of-the-art Transformer-based models in real production environments. 

\bibliographystyle{ACM-Reference-Format}
\balance
\bibliography{references}


\begin{thebibliography}{22}


\ifx \showCODEN    \undefined \def \showCODEN     #1{\unskip}     \fi
\ifx \showDOI      \undefined \def \showDOI       #1{#1}\fi
\ifx \showISBNx    \undefined \def \showISBNx     #1{\unskip}     \fi
\ifx \showISBNxiii \undefined \def \showISBNxiii  #1{\unskip}     \fi
\ifx \showISSN     \undefined \def \showISSN      #1{\unskip}     \fi
\ifx \showLCCN     \undefined \def \showLCCN      #1{\unskip}     \fi
\ifx \shownote     \undefined \def \shownote      #1{#1}          \fi
\ifx \showarticletitle \undefined \def \showarticletitle #1{#1}   \fi
\ifx \showURL      \undefined \def \showURL       {\relax}        \fi
\providecommand\bibfield[2]{#2}
\providecommand\bibinfo[2]{#2}
\providecommand\natexlab[1]{#1}
\providecommand\showeprint[2][]{arXiv:#2}

\bibitem[Ama(2023)]%
        {AmazonStatisticsUptoDate}
 \bibinfo{year}{2023}\natexlab{}.
\newblock \bibinfo{title}{Amazon {{Statistics}}: {{Up-to-Date Numbers
  Relevant}} for 2023-2024}.
\newblock
\newblock
\urldef\tempurl%
\url{https://amzscout.net/blog/amazon-statistics/}
\showURL{%
\tempurl}
\newblock
\shownote{[Online; accessed 16 January 2024]}.


\bibitem[Spo(2024)]%
        {SpotifyAnnoy2024}
 \bibinfo{year}{2024}\natexlab{}.
\newblock \bibinfo{title}{Spotify/Annoy}.
\newblock \bibinfo{howpublished}{Spotify}.
\newblock
\urldef\tempurl%
\url{https://github.com/spotify/annoy}
\showURL{%
\tempurl}


\bibitem[Abadi et~al\mbox{.}(2016)]%
        {abadiTensorFlowSystemLargeScale2016}
\bibfield{author}{\bibinfo{person}{Martin Abadi}, \bibinfo{person}{Paul
  Barham}, \bibinfo{person}{Jianmin Chen}, \bibinfo{person}{Zhifeng Chen},
  \bibinfo{person}{Andy Davis}, \bibinfo{person}{Jeffrey Dean},
  \bibinfo{person}{Matthieu Devin}, \bibinfo{person}{Sanjay Ghemawat},
  \bibinfo{person}{Geoffrey Irving}, \bibinfo{person}{Michael Isard},
  \bibinfo{person}{Manjunath Kudlur}, \bibinfo{person}{Josh Levenberg},
  \bibinfo{person}{Rajat Monga}, \bibinfo{person}{Sherry Moore},
  \bibinfo{person}{Derek~G. Murray}, \bibinfo{person}{Benoit Steiner},
  \bibinfo{person}{Paul Tucker}, \bibinfo{person}{Vijay Vasudevan},
  \bibinfo{person}{Pete Warden}, \bibinfo{person}{Martin Wicke},
  \bibinfo{person}{Yuan Yu}, {and} \bibinfo{person}{Xiaoqiang Zheng}.}
  \bibinfo{year}{2016}\natexlab{}.
\newblock \showarticletitle{TensorFlow: {{A System}} for Large-Scale {{Machine
  Learning}}}. In \bibinfo{booktitle}{\emph{12th {{USENIX Symposium}} on
  {{Operating Systems Design}} and {{Implementation}} ({{OSDI}} 16)}}.
  \bibinfo{pages}{265--283}.
\newblock


\bibitem[Chen et~al\mbox{.}(2020)]%
        {chenDifferentiableProductQuantization2020}
\bibfield{author}{\bibinfo{person}{Ting Chen}, \bibinfo{person}{Lala Li}, {and}
  \bibinfo{person}{Yizhou Sun}.} \bibinfo{year}{2020}\natexlab{}.
\newblock \showarticletitle{Differentiable {{Product Quantization}} for
  {{End-to-End Embedding Compression}}}. In \bibinfo{booktitle}{\emph{Proc.
  {{ICML}}}}.
\newblock


\bibitem[Cho et~al\mbox{.}(2011)]%
        {choFriendshipMobilityUser2011}
\bibfield{author}{\bibinfo{person}{Eunjoon Cho}, \bibinfo{person}{Seth~A.
  Myers}, {and} \bibinfo{person}{Jure Leskovec}.}
  \bibinfo{year}{2011}\natexlab{}.
\newblock \showarticletitle{Friendship and Mobility: User Movement in
  Location-Based Social Networks}. In \bibinfo{booktitle}{\emph{Proc.
  {{KDD}}}}. \bibinfo{pages}{1082--1090}.
\newblock


\bibitem[Goldenberg and Levin(2021)]%
        {goldenberg_bookingcom_2021}
\bibfield{author}{\bibinfo{person}{Dmitri Goldenberg} {and}
  \bibinfo{person}{Pavel Levin}.} \bibinfo{year}{2021}\natexlab{}.
\newblock \showarticletitle{Booking.com Multi-Destination Trips Dataset}. In
  \bibinfo{booktitle}{\emph{Proc. {SIGIR}}}. \bibinfo{pages}{2457--2462}.
\newblock


\bibitem[J{\'e}gou et~al\mbox{.}(2011)]%
        {jegouProductQuantizationNearest2011}
\bibfield{author}{\bibinfo{person}{Herve J{\'e}gou}, \bibinfo{person}{Matthijs
  Douze}, {and} \bibinfo{person}{Cordelia Schmid}.}
  \bibinfo{year}{2011}\natexlab{}.
\newblock \showarticletitle{Product {{Quantization}} for {{Nearest Neighbor
  Search}}}.
\newblock \bibinfo{journal}{\emph{IEEE Transactions on Pattern Analysis and
  Machine Intelligence}} \bibinfo{volume}{33}, \bibinfo{number}{1}
  (\bibinfo{year}{2011}), \bibinfo{pages}{117--128}.
\newblock


\bibitem[Johnson et~al\mbox{.}(2021)]%
        {FAISS}
\bibfield{author}{\bibinfo{person}{Jeff Johnson}, \bibinfo{person}{Matthijs
  Douze}, {and} \bibinfo{person}{Herv{\'e} J{\'e}gou}.}
  \bibinfo{year}{2021}\natexlab{}.
\newblock \showarticletitle{Billion-{{Scale Similarity Search}} with {{GPUs}}}.
\newblock \bibinfo{journal}{\emph{IEEE Transactions on Big Data}}
  \bibinfo{volume}{7}, \bibinfo{number}{3} (\bibinfo{year}{2021}),
  \bibinfo{pages}{535--547}.
\newblock
\showISSN{2332-7790}


\bibitem[Kang and McAuley(2018)]%
        {SASRec}
\bibfield{author}{\bibinfo{person}{Wang-Cheng Kang} {and}
  \bibinfo{person}{Julian McAuley}.} \bibinfo{year}{2018}\natexlab{}.
\newblock \showarticletitle{Self-{{Attentive Sequential Recommendation}}}. In
  \bibinfo{booktitle}{\emph{Proc. {{ICDM}}}}. \bibinfo{pages}{197--206}.
\newblock


\bibitem[Klenitskiy and Vasilev(2023)]%
        {klenitskiyTurningDrossGold2023}
\bibfield{author}{\bibinfo{person}{Anton Klenitskiy} {and}
  \bibinfo{person}{Alexey Vasilev}.} \bibinfo{year}{2023}\natexlab{}.
\newblock \showarticletitle{Turning {{Dross Into Gold Loss}}: Is {{BERT4Rec}}
  Really Better than {{SASRec}}?}. In \bibinfo{booktitle}{\emph{Proc. RecSys}}.
  \bibinfo{pages}{1120--1125}.
\newblock


\bibitem[Lian et~al\mbox{.}(2020)]%
        {lianLightRecMemorySearchEfficient2020}
\bibfield{author}{\bibinfo{person}{Defu Lian}, \bibinfo{person}{Haoyu Wang},
  \bibinfo{person}{Zheng Liu}, \bibinfo{person}{Jianxun Lian},
  \bibinfo{person}{Enhong Chen}, {and} \bibinfo{person}{Xing Xie}.}
  \bibinfo{year}{2020}\natexlab{}.
\newblock \showarticletitle{{{LightRec}}: {{A Memory}} and {{Search-Efficient
  Recommender System}}}. In \bibinfo{booktitle}{\emph{Proc. {{WWW}}}}.
  \bibinfo{pages}{695--705}.
\newblock


\bibitem[Liu et~al\mbox{.}(2024)]%
        {liuVectorQuantizationRecommender2024}
\bibfield{author}{\bibinfo{person}{Qijiong Liu}, \bibinfo{person}{Xiaoyu Dong},
  \bibinfo{person}{Jiaren Xiao}, \bibinfo{person}{Nuo Chen},
  \bibinfo{person}{Hengchang Hu}, \bibinfo{person}{Jieming Zhu},
  \bibinfo{person}{Chenxu Zhu}, \bibinfo{person}{Tetsuya Sakai}, {and}
  \bibinfo{person}{Xiao-Ming Wu}.} \bibinfo{year}{2024}\natexlab{}.
\newblock \bibinfo{title}{Vector {{Quantization}} for {{Recommender Systems}}:
  {{A Review}} and {{Outlook}}}.
\newblock
\newblock
\showeprint[arxiv]{2405.03110}~[cs]


\bibitem[Petrov and Macdonald(2025)]%
        {petrovRSSEffectiveEfficient2025}
\bibfield{author}{\bibinfo{person}{Aleksandr Petrov} {and}
  \bibinfo{person}{Craig Macdonald}.} \bibinfo{year}{2025}\natexlab{}.
\newblock \showarticletitle{{{RSS}}: {{Effective}} and {{Efficient Training}}
  for {{Sequential Recommendation Using Recency Sampling}}}.
\newblock \bibinfo{journal}{\emph{ACM Transactions on Recommender Systems}}
  \bibinfo{volume}{3}, \bibinfo{number}{1} (\bibinfo{year}{2025}),
  \bibinfo{pages}{1--32}.
\newblock
\showISSN{2770-6699}


\bibitem[Petrov and Macdonald(2022)]%
        {Bert4RecRepro}
\bibfield{author}{\bibinfo{person}{Aleksandr~V. Petrov} {and}
  \bibinfo{person}{Craig Macdonald}.} \bibinfo{year}{2022}\natexlab{}.
\newblock \showarticletitle{A {{Systematic Review}} and {{Replicability Study}}
  of {{BERT4Rec}} for {{Sequential Recommendation}}}. In
  \bibinfo{booktitle}{\emph{Proc. {{RecSys}}}}. \bibinfo{pages}{436--447}.
\newblock


\bibitem[Petrov and Macdonald(2023)]%
        {petrovGSASRecReducingOverconfidence2023}
\bibfield{author}{\bibinfo{person}{Aleksandr~V. Petrov} {and}
  \bibinfo{person}{Craig Macdonald}.} \bibinfo{year}{2023}\natexlab{}.
\newblock \showarticletitle{{{gSASRec}}: {{Reducing Overconfidence}} in
  {{Sequential Recommendation Trained}} with {{Negative Sampling}}}. In
  \bibinfo{booktitle}{\emph{Proc. {{RecSys}}}}. \bibinfo{pages}{116--128}.
\newblock


\bibitem[Petrov and Macdonald(2024)]%
        {petrovRecJPQTrainingLargeCatalogue2024}
\bibfield{author}{\bibinfo{person}{Aleksandr~V. Petrov} {and}
  \bibinfo{person}{Craig Macdonald}.} \bibinfo{year}{2024}\natexlab{}.
\newblock \showarticletitle{{{RecJPQ}}: {{Training Large-Catalogue Sequential
  Recommenders}}}. In \bibinfo{booktitle}{\emph{Proc. {{WSDM}}}}.
\newblock


\bibitem[Sun et~al\mbox{.}(2019)]%
        {BERT4Rec}
\bibfield{author}{\bibinfo{person}{Fei Sun}, \bibinfo{person}{Jun Liu},
  \bibinfo{person}{Jian Wu}, \bibinfo{person}{Changhua Pei},
  \bibinfo{person}{Xiao Lin}, \bibinfo{person}{Wenwu Ou}, {and}
  \bibinfo{person}{Peng Jiang}.} \bibinfo{year}{2019}\natexlab{}.
\newblock \showarticletitle{{{BERT4Rec}}: {{Sequential Recommendation}} with
  {{Bidirectional Encoder Representations}} from {{Transformer}}}. In
  \bibinfo{booktitle}{\emph{Proc. {{CIKM}}}}. \bibinfo{pages}{1441--1450}.
\newblock


\bibitem[Tonellotto et~al\mbox{.}(2018)]%
        {tonellottoEfficientQueryProcessing2018}
\bibfield{author}{\bibinfo{person}{Nicola Tonellotto}, \bibinfo{person}{Craig
  Macdonald}, {and} \bibinfo{person}{Iadh Ounis}.}
  \bibinfo{year}{2018}\natexlab{}.
\newblock \showarticletitle{Efficient {{Query Processing}} for {{Scalable Web
  Search}}}.
\newblock \bibinfo{journal}{\emph{Foundations and Trends{\textregistered} in
  Information Retrieval}} \bibinfo{volume}{12}, \bibinfo{number}{4-5}
  (\bibinfo{year}{2018}), \bibinfo{pages}{319--500}.
\newblock


\bibitem[Turtle and Flood(1995)]%
        {turtleQueryEvaluationStrategies1995b}
\bibfield{author}{\bibinfo{person}{Howard Turtle} {and} \bibinfo{person}{James
  Flood}.} \bibinfo{year}{1995}\natexlab{}.
\newblock \showarticletitle{Query Evaluation: {{Strategies}} and
  Optimizations}.
\newblock \bibinfo{journal}{\emph{Information Processing \& Management}}
  \bibinfo{volume}{31}, \bibinfo{number}{6} (\bibinfo{year}{1995}),
  \bibinfo{pages}{831--850}.
\newblock


\bibitem[Vaswani et~al\mbox{.}(2017)]%
        {Transformer}
\bibfield{author}{\bibinfo{person}{Ashish Vaswani}, \bibinfo{person}{Noam
  Shazeer}, \bibinfo{person}{Niki Parmar}, \bibinfo{person}{Jakob Uszkoreit},
  \bibinfo{person}{Llion Jones}, \bibinfo{person}{Aidan~N Gomez},
  \bibinfo{person}{{\L}ukasz Kaiser}, {and} \bibinfo{person}{Illia
  Polosukhin}.} \bibinfo{year}{2017}\natexlab{}.
\newblock \showarticletitle{Attention Is {{All}} You {{Need}}}. In
  \bibinfo{booktitle}{\emph{Proc. {{NeurIPS}}}}.
\newblock


\bibitem[Xia et~al\mbox{.}(2023)]%
        {xiaEfficientOnDeviceSessionBased2023}
\bibfield{author}{\bibinfo{person}{Xin Xia}, \bibinfo{person}{Junliang Yu},
  \bibinfo{person}{Qinyong Wang}, \bibinfo{person}{Chaoqun Yang},
  \bibinfo{person}{Nguyen Quoc~Viet Hung}, {and} \bibinfo{person}{Hongzhi
  Yin}.} \bibinfo{year}{2023}\natexlab{}.
\newblock \showarticletitle{Efficient {{On-Device Session-Based
  Recommendation}}}.
\newblock \bibinfo{journal}{\emph{ACM Transactions on Information Systems}}
  \bibinfo{volume}{41}, \bibinfo{number}{4} (\bibinfo{year}{2023}),
  \bibinfo{pages}{1--24}.
\newblock
\showISSN{1046-8188, 1558-2868}


\bibitem[Zhan et~al\mbox{.}(2021)]%
        {zhanJointlyOptimizingQuery2021}
\bibfield{author}{\bibinfo{person}{Jingtao Zhan}, \bibinfo{person}{Jiaxin Mao},
  \bibinfo{person}{Yiqun Liu}, \bibinfo{person}{Jiafeng Guo},
  \bibinfo{person}{Min Zhang}, {and} \bibinfo{person}{Shaoping Ma}.}
  \bibinfo{year}{2021}\natexlab{}.
\newblock \showarticletitle{Jointly {{Optimizing Query Encoder}} and {{Product
  Quantization}} to {{Improve Retrieval Performance}}}. In
  \bibinfo{booktitle}{\emph{Proc. {{CIKM}}}}. \bibinfo{pages}{2487--2496}.
\newblock


\end{thebibliography}
\clearpage

\end{document}